\documentclass{pasj00} 
\begin{document}
 
\def\kms{km s$^{-1}$} \def\deg{$^\circ$ } \def\Deg{^\circ} 
\def\Msun{M_{\odot \hskip-5.2pt \bullet}} 
\def\v{\vskip 2mm} \def\bc{\begin{center}} 
\def\ec{\end{center}} \def\be{\begin{equation}}
\def\ee{\end{equation}}  \def\A{ Alfv{\'e}n } 

\title{The Primordial Origin Model of Magnetic Fields in Spiral Galaxies}
\author{Yoshiaki {\sc Sofue}$^{1,2}$, Mami {\sc Machida}$^{3,4}$, and Takahiro {\sc Kudoh}$^{3}$}
\affil{1. Dept. Physics, Meisei University, Hino, Tokyo 191-8506, \\
2. Inst. of Astronomy, University of Tokyo, Mitaka, Tokyo 181-8588, \\
3. National Astronomical Observatory of Japan, Mitaka, Tokyo 181-8588,\\
4. Dept. of Physics, Kyushu University, Higashi-ku, Fukuoka 812-8581 ,\\ 
{\it E-mail: sofue@ioa.s.u-tokyo.ac.jp} }


\maketitle

\begin{abstract}  
We propose a primordial-origin model for the composite configurations of global magnetic fields in spiral galaxies. We show that uniform tilted magnetic field wound up into a rotating disk galaxy can evolve into composite magnetic configurations comprising bisymmetric spiral (S=BSS), axisymmetric spiral (A=ASS), plane-reversed spiral (PR), and/or ring (R) fields in the disk, and vertical (V) fields in the center. By MHD simulations we show that these composite galactic fields are indeed created from weak primordial uniform field, and that the different configurations can co-exist in the same galaxy. We show that spiral fields trigger the growth of two-armed gaseous arms. The centrally accumulated vertical fields are twisted and produce jet toward the halo. We find that the more vertical was the initial uniform field, the stronger is the formed magnetic field in the galactic disk.
\end{abstract}

\section{Introduction}

\begin{figure}\bc 
\includegraphics[width=8cm]{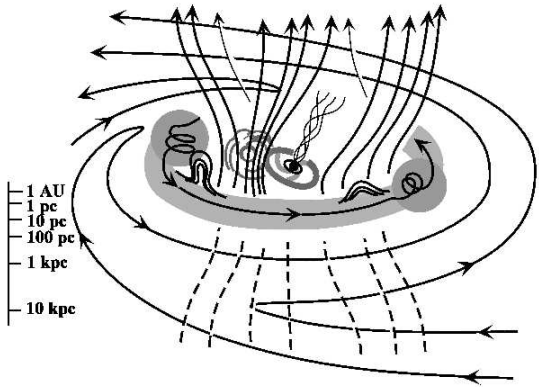}\ec
\caption{Schematic illustration of magnetic fields in spiral galaxies, including our Galaxy. }  
\label{fig1} 
\end{figure}

Global magnetic fields in disk galaxies are observed to show either bisymmetric spiral (BSS, hereafter S), ring (R) or axisymmetric (ASS, hereafter A) configuration (e.g., Sofue et al 1987). Figure \ref{fig1} illustrates the observed magnetic configurations. Large-scale S fields on the order of $\sim 1~\mu$G are observed in many spiral galaxies including the Milky Way (Sofue et al. 1986; Sofue et al. 1983; Han et al. 1998; Krause et al. 1989, 1998). An R field is observed in M31 (Berkhuijsen et al. 2003), while some galaxies such as IC 342 show A fields (Krause et al. 1989). In addition to the S configuration, the local field'" in the Milky Way are observed to show reversal of field direction with respect to the galactic plane from positive to negative galactic latitudes (Sofue and Fujimoto 1983; Taylor et al. 2009). We call such configuration the galactic-plane-reversed (GPR) field.

Edge-on galaxies often indicate V fields in their halos (Brandenburg et al. 1993; Krause et al. 2008). The central regions of spiral galaxies exhibit V fields as observed for M31 (Berkhuijsen et al. 2003) and M81 (Krause et al. 1989). The Milky Way's center indicates prominent V fields as strong as $\sim 1$ mG (Yusef-Zadeh et al. 1984; Tsuboi et al. 1986; Sofue et al. 1986; LaRosa et al. 2005), where the magnetic energy density (pressure) much exceeds the thermal and turbulent energy densities of the interstellar gas, showing that the energy-density equipartition does not hold. The field strength is maintained steady due to amplification by dynamo mechanism (Parker 1979) and annihilation by magnetic diffusion (e.g. Fujimoto and Sawa 1987).  

We emphasize that the large-scale magnetic configurations in spiral galaxies are a mixture of different types of configurations, and that they co-exist in one galaxy. Figure \ref{fig1} illustrates the global field configuration in the Galaxy, which may generally apply to spiral galaxies. The origin of such composite topology has remained as a mystery. In this paper we try to solve this problem, and to explain the observed configurations as the fossil of primordial magnetic field when the galaxies were formed. We unify the basic topologies - S, A, RA, R and V - into a common origin.

Magnetohydrodynamic (MHD) simulations of accretion disks have been extensively performed to study the dynamics and jet formation in magnetized galactic disks and AGN torus around massive black holes (Kudoh et al. 1998; Matsumoto 1999;  Machida et al. 2000; Nishikori et al. 2006). Nishikori et al. (2006) traced the three dimensional MHD evolution of a galactic disk with a weak azimuthal field, showing that the fields are amplified by the differential rotation as well as by the magneto-rotational instabilities (MRI: Balbus and Hawley 1991; Hawley and Balbus 1991). They showed that a tightly wound spiral configuration superposed by random fields is formed. Wang and Abel (2009) have recently performed MHD simulations of formation of a disk galaxy and magnetization of interstellar medium. They have shown that a weak uniform field perpendicular to the disk is amplified to create micro G interstellar fields during the disk formation, which affects the structure formation of ISM and star formation. 

In this paper, we also try to confirm that our proposed scenario for the global magnetic configurations indeed works in a disk galaxy by three dimensional MHD simulations. We focus on the global topology of magnetic lines of force, and show that a unified initial field can evolve into various magnetic configurations as observed in spiral galaxies. The present paper would be, therefore, complimentary to the above cited papers, in which more detailed ISM physics and jet formations are discussed.

\section{Primordial Origin Hypothesis } 

\subsection{S (BSS)  Field}

\begin{figure}\bc
\includegraphics[width=8cm]{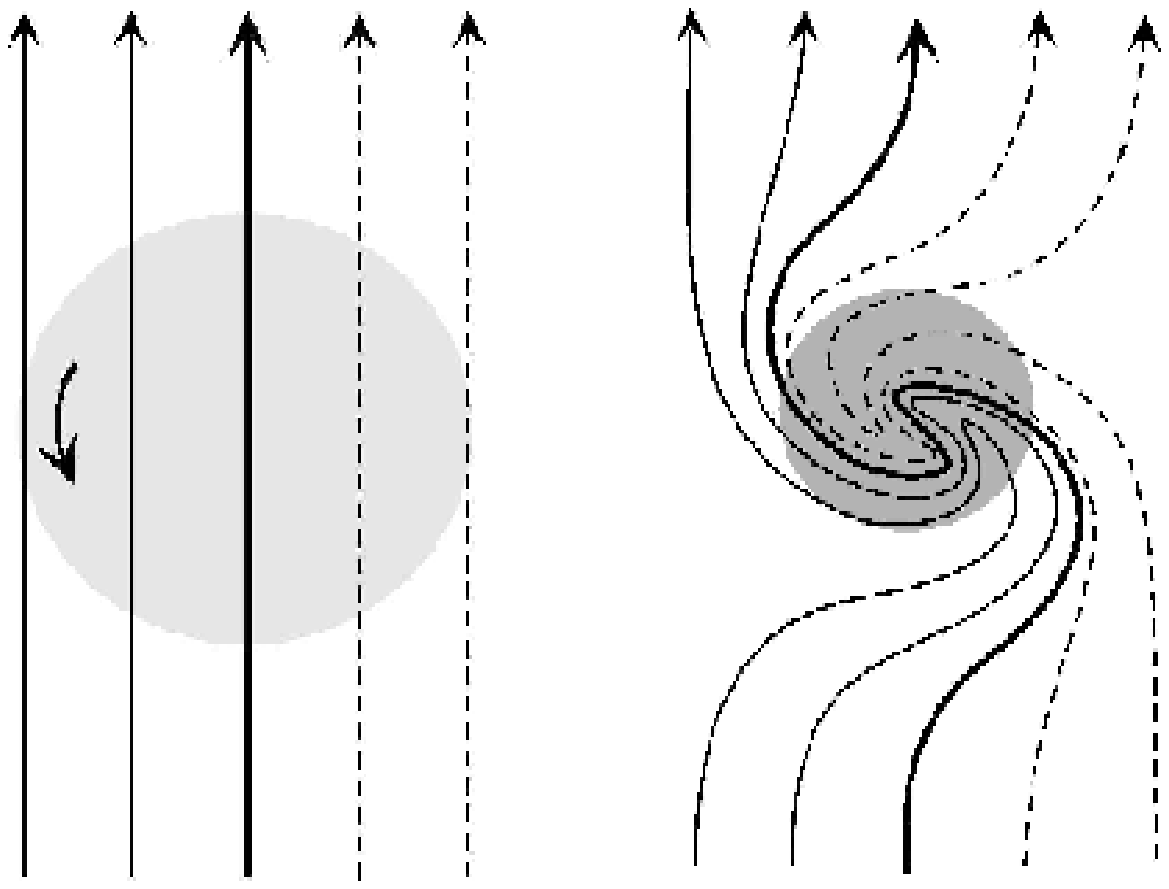}
\includegraphics[width=8cm]{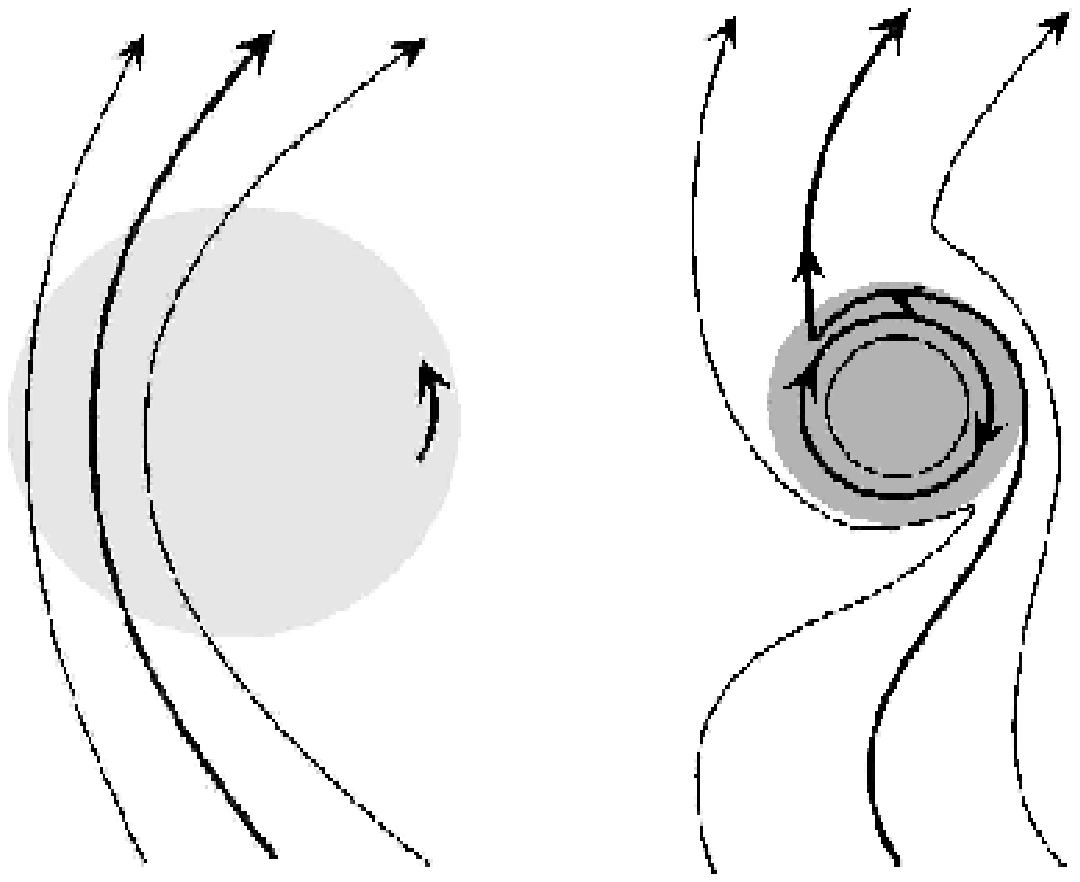}
\includegraphics[width=8cm]{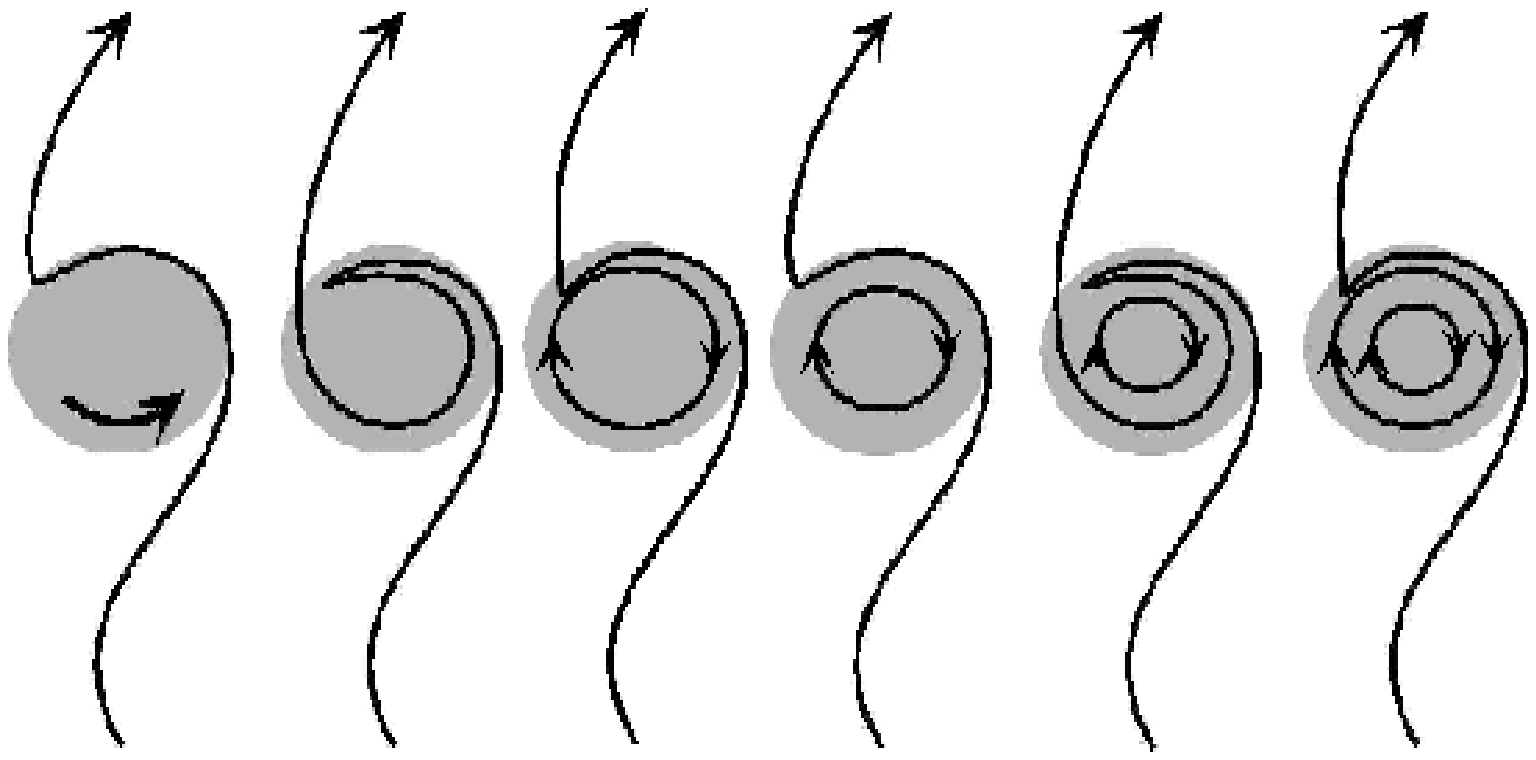}\ec
\caption{Cosmological origin of S field from a uniform field, and of R field and its amplification.}
\label{fig2} 
\end{figure}

It has been suggested that the S configuration may be created by a rotating disk galaxy, if a larger scale primordial magnetic field of cosmological origin were wound up into the rotating disk (Sofue et al. 1986, 1987).  Figure \ref{fig2} illustrates this idea: A large scale uniform magnetic field parallel to the galactic plane is wound up with the rotating disk. When the disk formation is completed, the S field is maintained with a constant pitch angle by magnetic reconnection near the neutral sheets. The field reconnection occurs in a time scale $t$ determined by $t=\lambda/V_{\rm A}$, which we call the \A time. Here the reversal scale $\lambda$ of the field may be taken as the turbulent cloud scale between neighboring spiral arms on the order of $\sim 100$ pc. The \A speed $V_{\rm A}$ is on the order of $\sim 6$ \kms for $B\sim 3 \mu$G and $\rho\sim 1$ atoms cm$^{-3}$. Then, we have $t\sim 2\times10^7$ years, which is sufficiently shorter than the rotation period of the ga
 laxy.

\begin{figure}\bc   
\includegraphics[width=8cm]{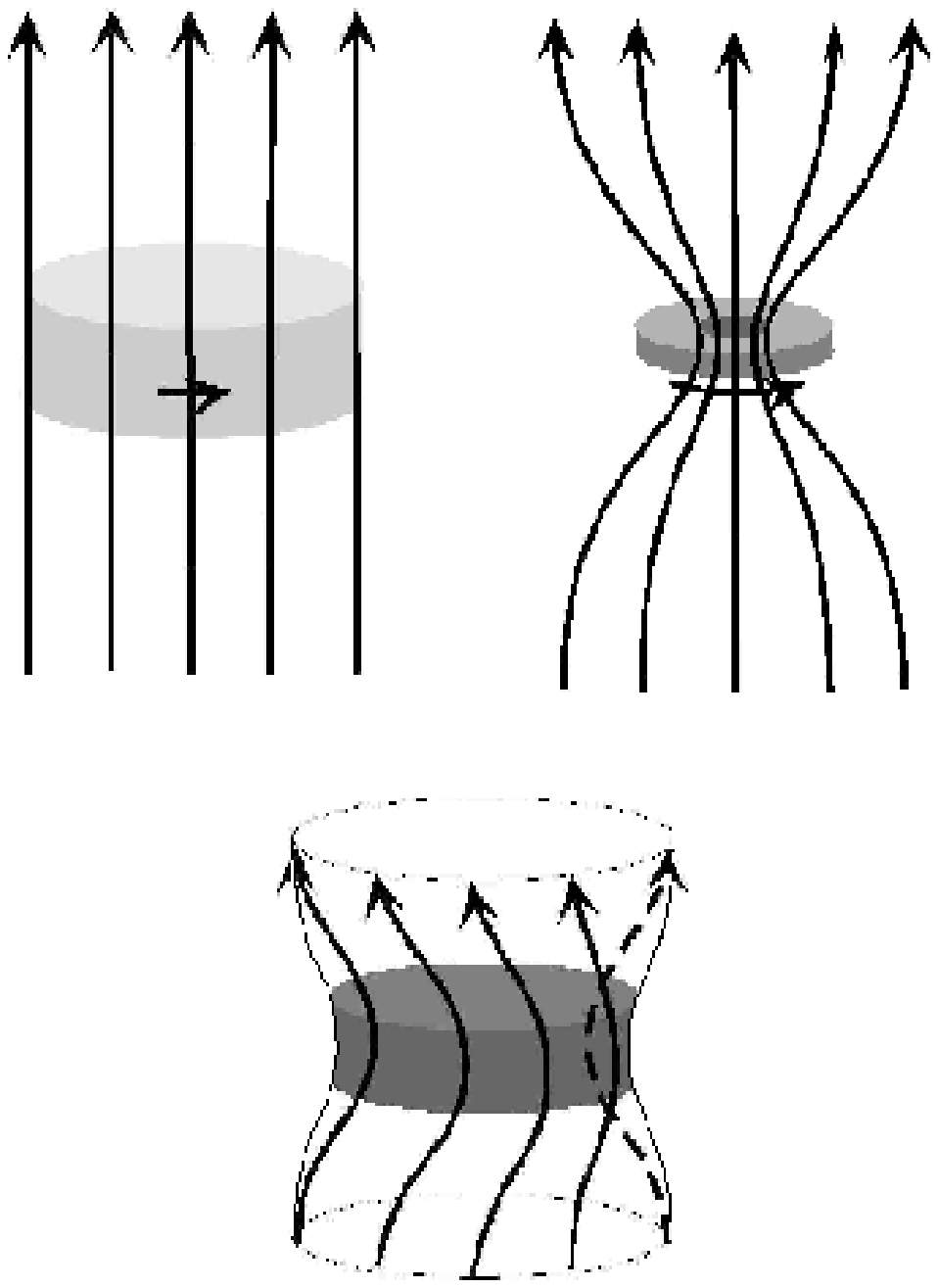} 
\includegraphics[width=5cm]{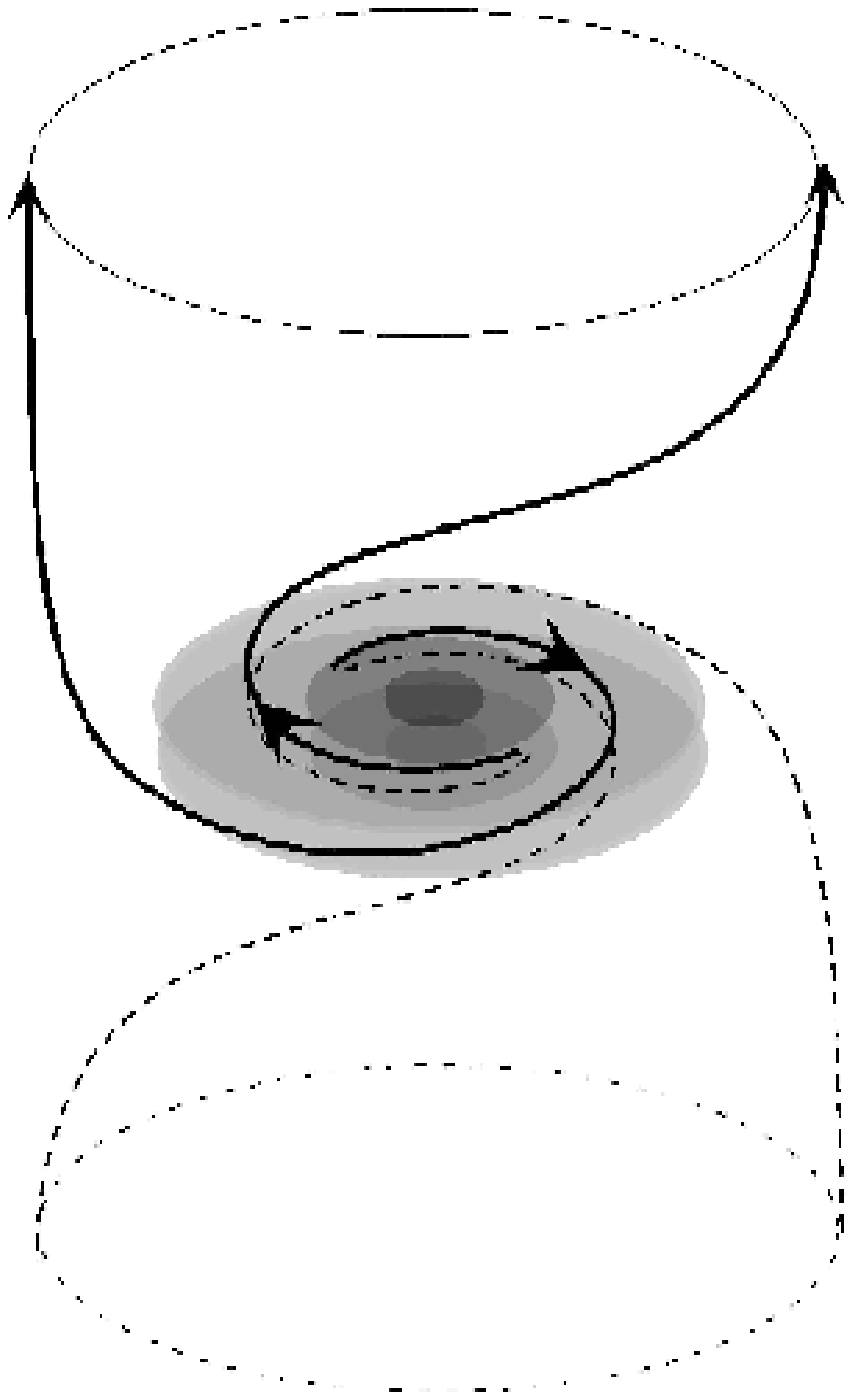}
\ec
\caption{Cosmological origin of V, A, and PR fields.}
\label{fig3} 
\end{figure}     

\subsection{R (Ring)  Field}

It may happen that the primordial magnetic field before galaxy formation was not uniform or lopsided such as due to proper motion of the primordial galaxy. Then, a part of the amplified S field will be reconnected in the galactic disk, and creates an R field inside a certain radius, as illustrated in Fig. \ref{fig2}. Once a ring field is created, it will shrink to attain smaller radius due to the magnetic tension and disk gas accretion toward the center. Accordingly, the bisymmetric component, left over in the outer disk, will be wound again by the differential rotation of the galaxy, and another reconnection takes place, creating a second ring. Thus, the field strength of the ring component increases, as the galaxy rotates.

\subsection{A (ASS) Field}

Some galaxies show axi-symmetric (A) configuration, where the field direction is always away from or toward the center, as if there exists a magnetic monopole in the center. Such field configuration is interpreted as due to locally enhanced polarized emission inside the disk by sheared magnetic loops, each of which is closed between the disk and halo. A more simple explanation of the A configuration is a primordial-origin from a vertical uniform field as shown in Fig. \ref{fig3}. If a vertical primordial field was wound up by the galactic disk, it is twisted tightly near the galactic disk. The upper parts of the wound field with respect to the galactic plane have spiral configuration with their directions away from or toward the galactic center, and in the lower half of the disk the field directions are reversed. Since the radio emission is strongest near the galactic plane, the Faraday rotation takes place for the emissions transferring the upper half of the magnetized disk,
  and hence, the observed rotation measure (RM) apparently indicates A configuration. 

\subsection{PR (Galactic Plane-Reversed) Field}

The distribution on the sky of Faraday rotation measure (RM) of quasars is often used to diagonyse the magnetic field configuration in the Milky Way (Sofue and Fujimoto 1983; Taylor et al. 2009). The observed longitudinal variation of RM, reversing from one arm direction to the next arm, is well explained by S type field. The RM distribution also shows large-scale latitudinal variation, showing large-scale reversal of field direction from positive to negative longitudes in the local arm, e.g. in the direction of $l\sim 90\Deg $ (Sofue and Fujimot 1983; Taylor et al. 2009). This indicates that the field direction is reversing in the upper and lower regions of the disk plane. Such galactic plane-reversed (PR) field may be related to the A field, as discussed below. The PR configuration may be created at the same time when an A field is created from a V field, as illustrated in figure \ref{fig3}.

\subsection{V (Vertical) Field}

The strong vertical field in the Galactic Center has been a mystery. One possible explanation is to attribute it to the fossil of primordial vertical field trapped to the galactic disk (Fig. \ref{fig3}). Suppose that a forming galactic disk was penetrated by a uniform magnetic field perpendicular to the disk plane. According to the contraction of the galactic gas disk, the field is accumulated to the central disk. The field lines cannot escape across the disk: The diffusion time due to anomalous diffusion defined by the \A time calculated for a length scale $\lambda \sim R \sim 3$ kpc for a disk galaxy is $\sim 5 \times 10^8$ years, much larger than the rotation period. The vertical field is twisted by the disk rotation, and causes angular momentum loss, which accelerates the disk contraction and amplify the V field. Thus, a strong V field is created in the nuclear gas disk, where the magnetic energy density becomes comparable to the kinetic energy density of gas due to the g
 alactic rotation. In fact the V field  of a few mG in the Galactic Center has energy density as high as $E_{\rm m}\sim 10^{-6}$ erg cm$^{-3}$, which is close to the kinetic energy density of the molecular gas by galactic rotation $\sim 100$ pc of $E_{\rm k} \sim 10^{-6}$ erg cm$^{-3}$, and is much larger than the thermal energy density $E_{\rm t}\sim 10^{-10}$ erg cm$^{-3}$.

\subsection{Unified Scenario for Primordial Origin} 

\begin{figure}\bc
\includegraphics[width=7cm]{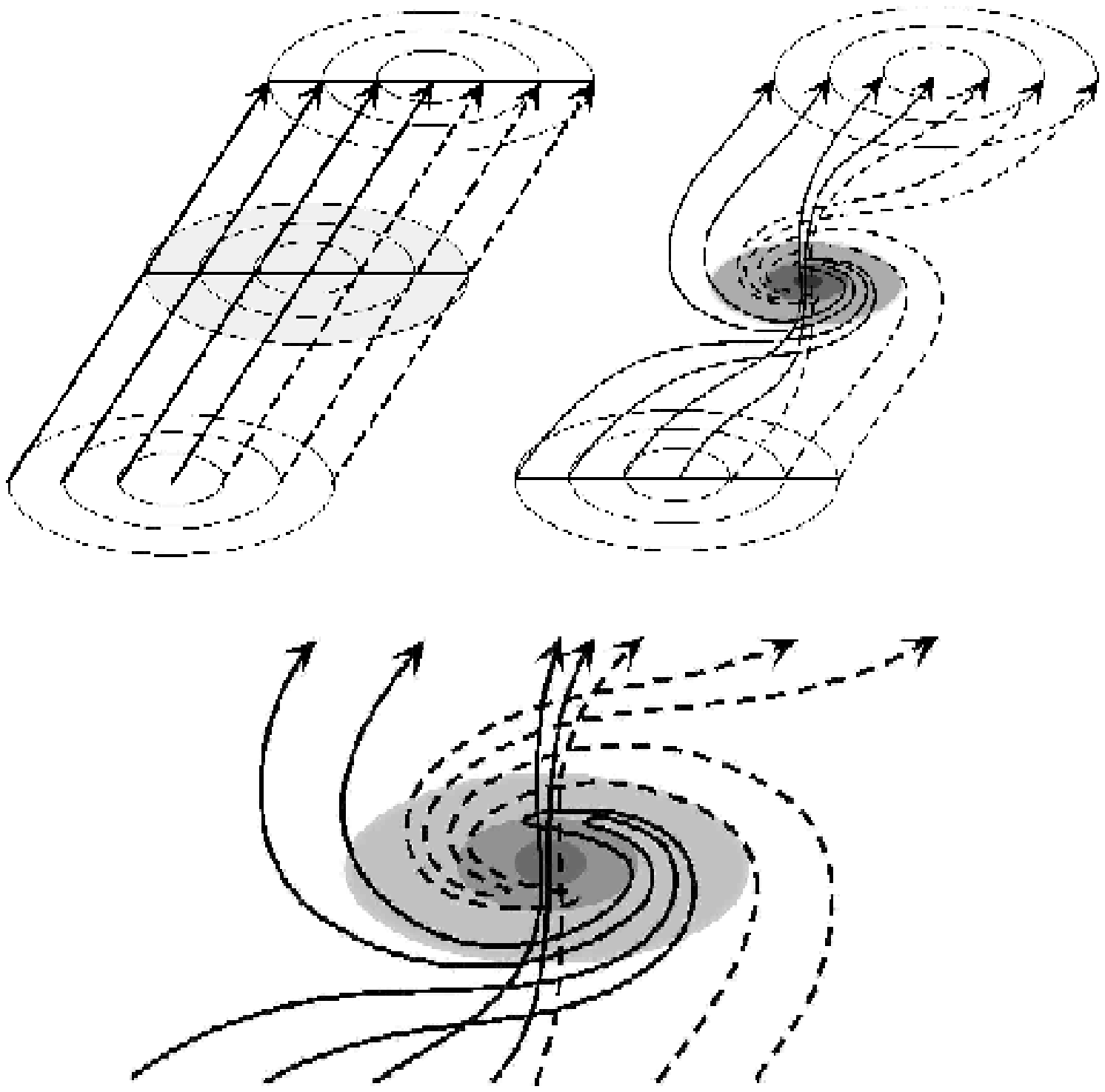}  
\includegraphics[width=7cm]{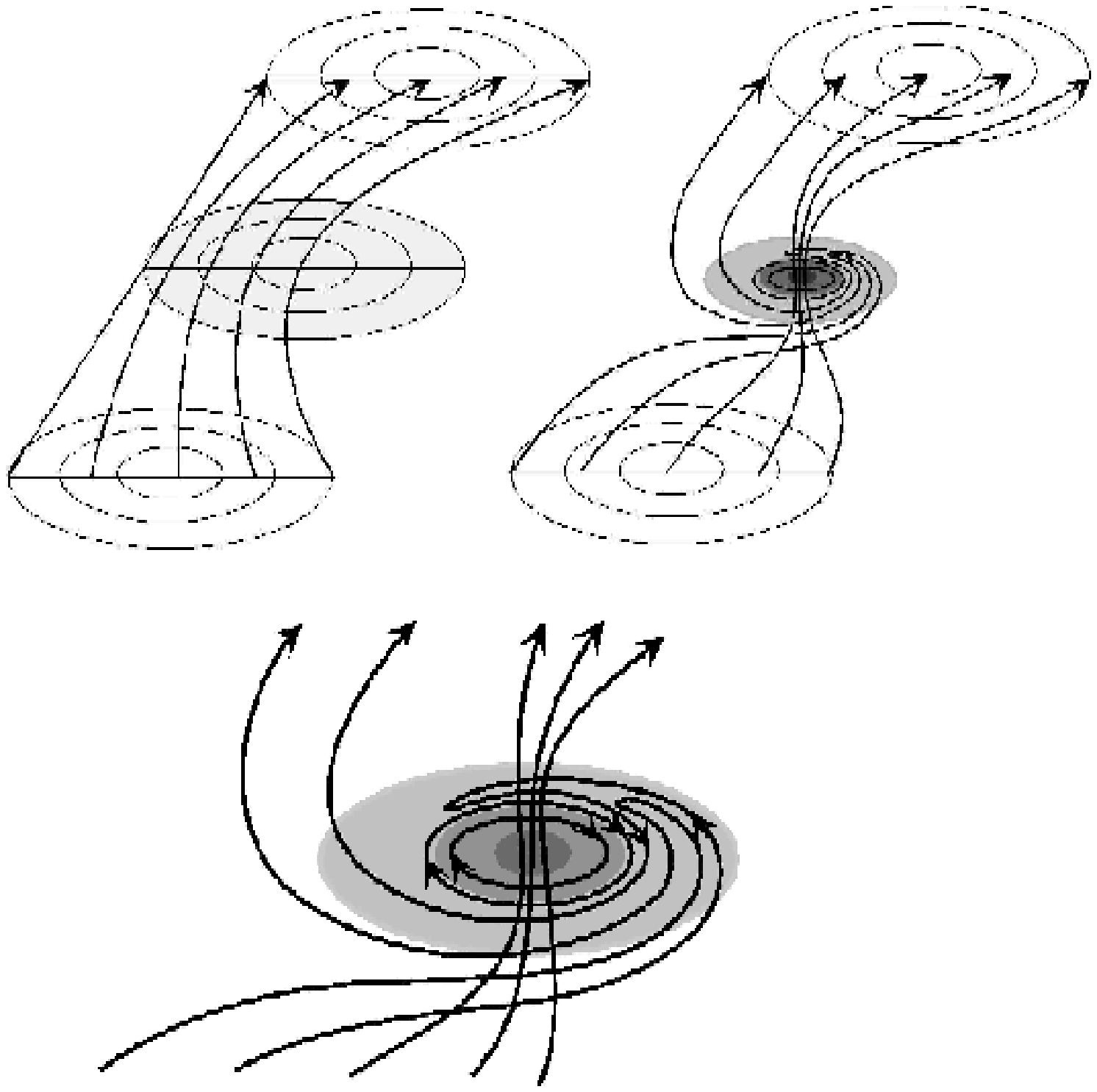} 
\ec
\caption{Origin of S, R and V fields from uniform (top) and lopsided fields (bottom).} 
\label{fig4}  
 \end{figure}

We, here, encounter a serious question why some or all of the different topologies, S (BSS), A (ASS), R (ring), PR (plane-reversed), and V (vertical), are observed simultaneously in one galaxy. In order to answer this question, we propose a unified scenario of primordial magnetic field origin during the disk formation of a galaxy. We consider a tilted original magnetic field penetrating a forming galactic disk, as illustrated in figure \ref{fig4}. As the lines of force are wound up by the contracting gas disk, the strength of vertical component is amplified simply obeying $B_{\rm v} \sim B_{\rm 0v} (R/R_0)^{-2}$, where $R$ is the radius and $R_0$ is the original radius within which the original field was enclosed. When $B_{\rm v}$ reaches a critical strength, the tension of the field will become strong enough so that the toroidal (spiral) field is slipped off from the disk plane to catch up the original tilted off-plane field. The amplified vertical field is further accumulat
 ed to produce a strong V field in the central disk. 

Thus, the outer field is wound up to form an S field in the outer disk, and a V field in the central region. There appears a composite field orientation: V in the center, and S field in the disk. If there was some anisotropy superposed on the tilted uniform field, a part of the S component in the disk will be reconnected to produce an R field. This will result in a composite magnetic structure composed of the three configurations: V in the center, R in the inner disk, and S field in the outer disk. The observed asymmetry of the rotation measures with respect to the galactic plane (Sofue and Fujimoto 1983) may be explained by an A component superposed on these configurations.    

If the original field was perpendicular to the disk, or the tilt angle was large enough, the wound-up field evolves in to spiral configuration with its direction always inward or outward in one side of the galactic disk. It shows the same structure in the other side of the disk, but the field direction is opposite, e.g. outward or inward. Thus, the field configuration is ASS type in one side, and the same but reversed in the other side of the disk. If the field configuration in the solar vicinity has such structure, we may explain the large-scale field reversal observed in the RM distribution on the sky. Also, linear polarization observations of radio emission from such a disk with a plane-reversed ASS field will be observed to show a global A field, since the Faraday rotation measure is more weighted by the ISM and field in the nearer side to the observer.

\section{MHD Simulation}

\subsection{The Method}
 
\def\nabra{\bigtriangledown}
\def\d{\partial}

In order to clarify if the here proposed primordial-origin scenario for the galactic magnetism indeed works during galaxy formation, we performed three-dimensional magneto-hydro dynamical (MHD) numerical simulations. The basic equations to be solved are as follows.
\be 
{\d \rho \over \d t}+  \nabra \cdot (\rho  {\bf v}) = 0,
\ee
\be
\rho\left[{\d{\bf v} \over \d t} + {\bf \it v}\cdot\nabra \right] = - \nabra P -\rho\nabra\phi +{{\bf j}\times{\bf  B} \over c},
\ee
\be
{\d {\bf B} \over \d t}=\nabra \times ({\bf v}\times{\bf B})-{4 \pi \over c} \eta_0 {\bf j}, 
\ee
\be
\rho T {dS \over dt}=0,
\ee 
where $\rho,~ P,~ \phi,~ {\bf v},~ {\bf B},~ {\bf j}=c \nabra \times {\bf B}/4 \pi,~ T,~ \eta_0 $ and $S$ 
are the density, pressure, gravitational potential, velocity, magnetic field, current density, temperature, magnetic diffusivity and specific entropy, respectively. We solve these equations using the modified Lax-Wendroff numerical method with the cylindrical 
coordinate $(r, \phi, z)$ (Matsumoto 1999; Nishikori et al. 2006). 
We here report only the essential points of the simulations in order to confirm  the reality of the present model. Detailed description of the numerical simulations will be given in a separate paper.  

We consider a gas disk rotating in a fixed gravitational potential of Miyamoto and Nagai (1975), which approximately represent the observed rotation curve of the Galaxy and is expressed by
\be
\phi(r,z)=-\Sigma_{i=1}^2 {GM_i \over [r^2+\{a_i+(z^2+b_i^2)^{1/2}\}^2]^{1/2}},
\ee
where the parameters $a_i$ and $b_i$ are given in table \ref{tab-MN}. 

\begin{table}
\caption{Parameters for the Miyamoto-Nagai potential.} 
\begin{center}
\begin{tabular}{lllll}
\hline\hline  \\  
Component & $i$ & $a_i$ (kpc) & $b_i$ (kpc) & $M (\Msun)$   \\
\\
\hline 
\\
Bulge & 1 & 0.0 & 0.495& $2.05 \times 10^{10}$ \\
Disk &2 & 7.258 & 0.520 & $2.547 \times 10^{11}$ \\ 
\\
\hline
\end{tabular} \\
\end{center} 
\label{tab-MN}
\end{table}
 
\begin{figure} \bc   
\includegraphics[width=8cm]{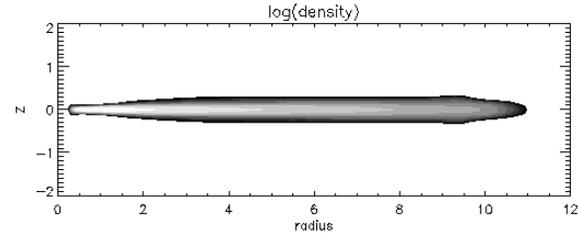} \ec
\caption{Initial gas density profile in the $(r,z)$ plane.}  
\label{fig5}  
\end{figure}

As the initial condition we assume that the disk is penetrated by a uniform intergalactic magnetic field. 
The inclination of the field lines to the rotation axis ($z$ direction), $\theta$, is taken to be 0\deg (perpendicular to disk),  45\deg, 80\deg and 90\deg (parallel). 
The interstellar gas is put freely, rotating in the potential with the centrifugal force being balanced with the gravity. The gas density is assumed to be exponential in the radial direction with a scale radius 7.26 kpc. 
The inner and the outer edges of the disk are 0.4 kpc and 10 kpc, respectively. 
The disk gas is given an equivalent temperature of $10^4$ K and a density of $\sim 1{\rm ~H ~cm}^{-3}$, so that the scale thickness of the disk is $H\sim 100$ pc. The intergalactic gas is assumed to have temperature of $10^6$ K. 
Figure \ref{fig5} shows the initial density profile in the $r$ and $z$ direction.
 
The grid numbers in the directions of radius, azimuthal angle, and height from the galactic plane, were 301, 66 and 337, respectively. 
The grid spacing's in the radius was 0.01 kpc up to 1 kpc, and it is stretched for $r>1$ kpc.
The spacing's in height direction is 0.002 kpc till 0.15 kpc, and it is stretched for $z>0.15$ kpc. The maximum grid spacing for $r$ and $z$ is 0.1kpc. In the azimuthal direction, it is uniform of 0.098 radian. 
The magnetic Reynolds number was taken to be $R_{\rm m0}=v_0r_0/\eta_0=2000$, where $\eta_0$ is the magnetic diffusivity. The corresponding magnetic diffusivity is $\eta_0 \simeq 3 \times 10^{25} {\rm cm}^2{\rm s}^{-1}$, where we assume 
the turbulent diffusivity (e.g., Parker 1979).
The initial plasma $\beta$ value at $r=r_0=1 $ kpc in the galactic plane was 100.  

The initial uniform field strength was given to be $\sim$ 0.6 $\mu$ G.  
The unit time of this simulation is the inverse of angular velocity, 
$t_0=r_0/v_0=4.7$ My, at $r_0=1$ kpc and $v_0=(GM_0/r_0)^{1/2}=207$ \kms for $M_0=10^{10}\Msun$. 
In order to mimic the galactic disk formation, the accumulated central gas is assumed to be transformed to stars, so that the gas density within central 200pc region is kept always low enough. 

\begin{table}
\caption{Units adopted in this paper.} 
\begin{center}
\begin{tabular}{lll}
\hline\hline  \\  
Physical quantity & Symbol & Numerical unit   \\
\\
\hline 
\\
Length & $r_0$   &  1 kpc \\
Velocity & $v_0$ &  207 \kms \\
Time & $t_0$ & $4.7\times 10^6$ yr \\  
Density & $\rho_0$ & $ 1.6 \times 10^{-24}$ g cm$^{-3}$ \\  
Magnetic field & $B_0=\sqrt{\rho_0v_0^2}$ & 26 $\mu$G \\ 
\\
\hline
\end{tabular} \\
\end{center} 
\label{tab-unit}
\end{table}

\subsection{Results of MHD simulations}

The result of MHD simulations is shown in Fig. \ref{fig6}. As the gas disk rotates differentially, the magnetic lines of force are twisted and wound up with the disk. When the field lines are wound tightly enough, they are reconnected at the neutral sheet within the magnetic diffusion time, which is on the order of ten times the crossing time of \A wave across the disk thickness $\lambda\sim d$. The diffusion time is longer in the outer disk, while it is shorter in the central region, because the disk thickness is larger in the outer disk.  
  
Figures \ref{fig8}, \ref{fig9} and \ref{fig10} show the
distributions of logarithm of the gas density and magnetic energy density for the case of $\theta=0\Deg,~45\Deg,$ and $90\Deg$,
respectively. The gas and magnetic energy densities are averaged around the
equatorial plane ($|z|<0.02$).  

\subsection{Evolution of vertical field}

\begin{figure*} \bc 
\includegraphics[width=12cm]{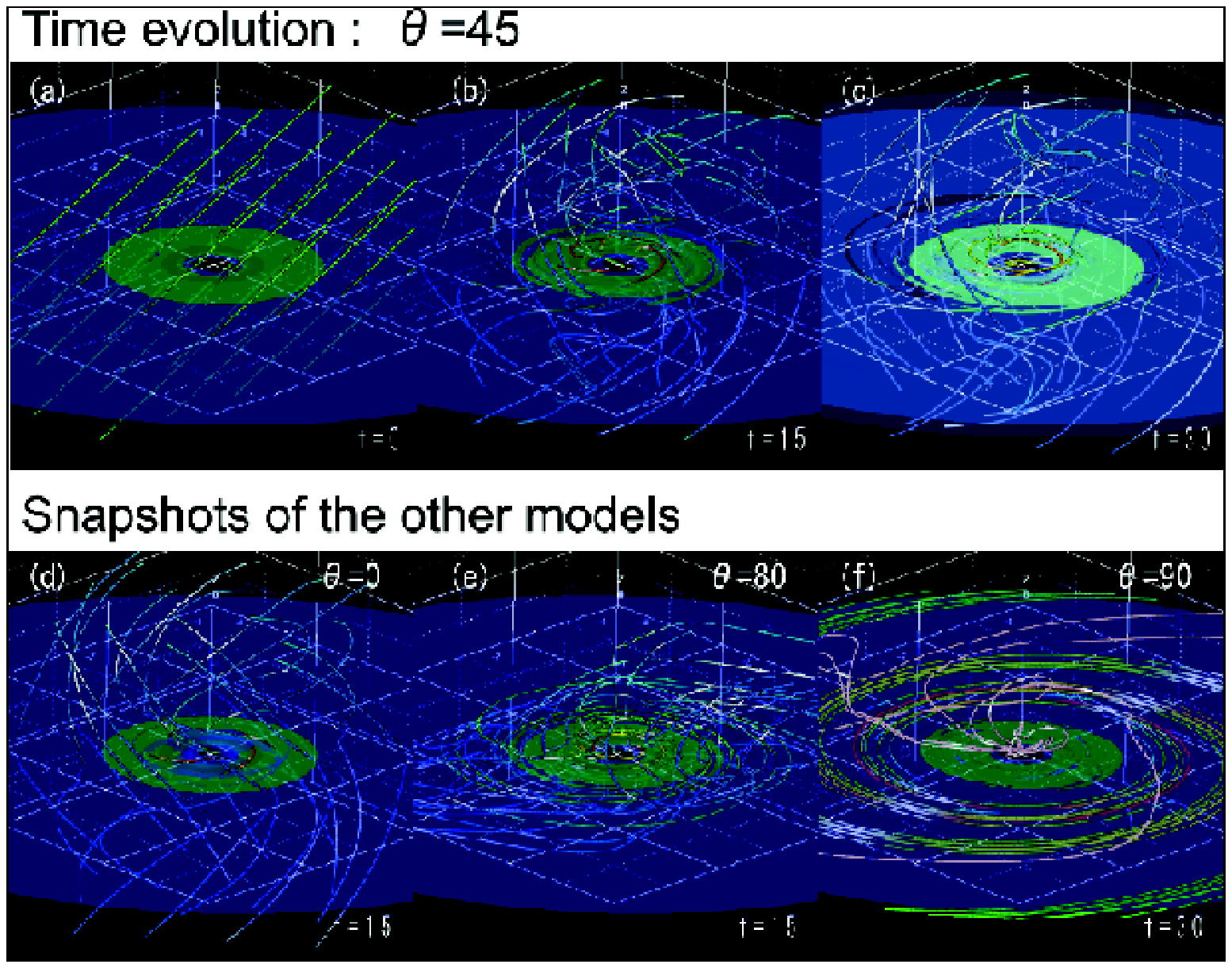}  \ec 
\caption{MHD simulation of the primordial origin model of
    magnetic field in a disk galaxy. Curves show the magnetic lines of
    force. Blue and green surface show the iso-surface of the density at $\rho
    = 0.1$ and $0.28$, respectively. (a)-(c): Evolution of field lines
    originally inclined by $\theta=45\Deg$ to the rotation axis at $t=
    0,~14.5$ and 30.0 $t_0$ with $t_0=4.7$ My. (d)-(f): Field lines for
    $\theta=0\Deg,~80\Deg$ at $14.5t_0$ and $90\Deg$ at $t=30.0t_0$. 
    Note that S or A, and V fields are formed in the same galaxy. }  
\label{fig6}   
\bc   
\includegraphics[width=6cm]{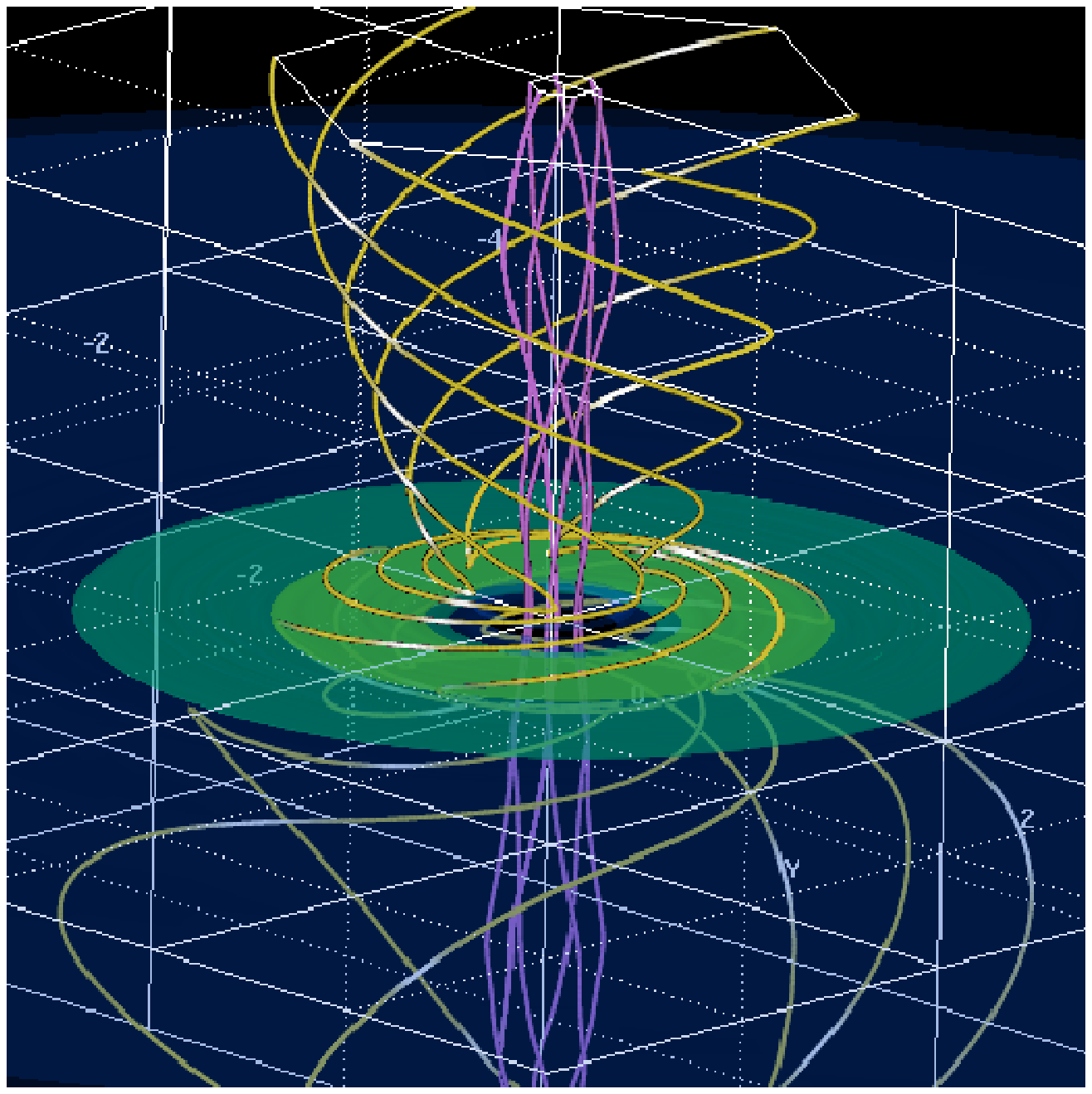} \ec 
\caption{ Central region of Fig. \ref{fig6}d.}  
\label{fig7}  
\end{figure*} 
 
In the case of $\theta=0\Deg$, the vertical field lines that penetrated the outer disk are wound up to produce spiral field whose direction reverses at the galactic plane. In the upper side of the disk, the field lines point inwards (or outwards) at any azimuthal angle, showing A configuration, whose lower half shows opposite field direction. In the central region of the disk, the initial field lines are twisted to create a helical V field. The global field configuration becomes, thus, a composite of outer A and inner V fields as shown in Fig. \ref{fig7}. According to the twisting, the upper and lower inversely directing components are reconnected near the disk plane, which results in accumulation of the vertical field lines toward the central region. The accumulated vertical fields are further twisted in the center, creating stronger vertical field. 
 
Thus a very strong vertical field is created in the central region. 
In the present computations, the elapsed time corresponds to a few hundred million years, or one galactic rotation time near the solar orbit. However, the galactic rotation within the central 1 kpc is on the order of ten million years, so that the disk rotates almost ten times. In such rotation periods, the vertical component is strongly accumulated, and strength is amplified by a factor of $(r/r_0)^2$, where $r$ is the final radius of the vertical field tube, and $r_0$ is its initial value. 
If $r_0 \sim 10$ kpc, and the $r\sim 1$ kpc, the initial field is amplified by
a factor of $10^2$. 
Hence, even if the initial field was very weak, very strong central vertical
field is created. 

 According to the accumulation and twisting, the field strength near the center is significantly amplified, and the magnetic pressure gradient perpendicular to the disk becomes strong enough to accelerate the gas toward the halo. This causes vertical outflow of gas as well as the loss of angular momentum of the disk by magnetic braking, accelerating the contraction of the galactic disk (Uchida et al. 1986).  

\begin{figure}
\bc
\includegraphics[width=8cm]{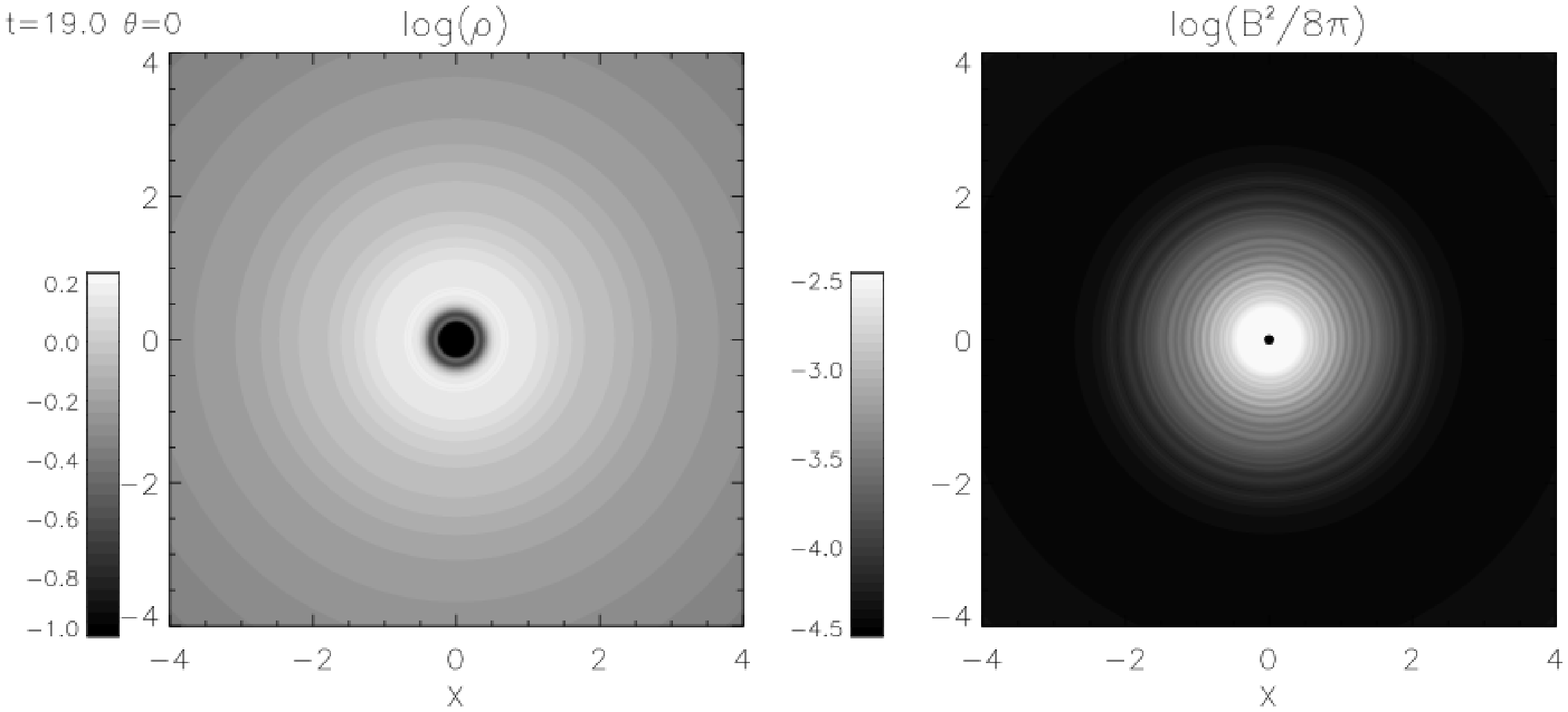} 
\ec
\caption{Gas density log $\rho$ (left) and magnetic energy density log $B^2/8
  \pi$ (right) for the case of $\theta =0$. Here, $B$ and $\rho$ are
  normalized by $B_0=26~\mu$G and  $\rho_0=1~ {\rm H~cm}^{-3}$. The gas
  density is maximum near the neutral sheet, where the field strength is minimum.}  
\label{fig8} 
\bc
\includegraphics[width=8cm]{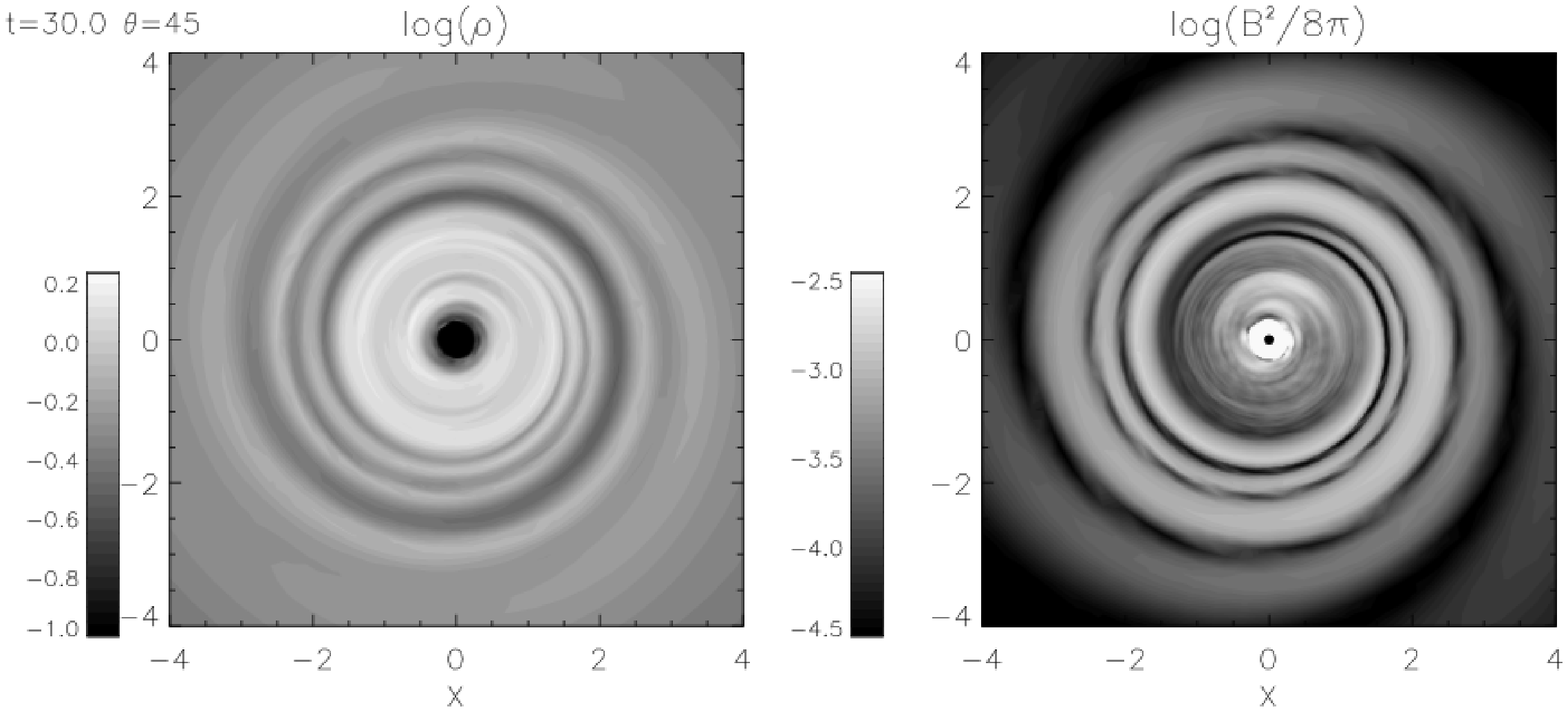}
\ec
\caption{Same as fig.\ref{fig8}, but for the case of $\theta=45\Deg$.}
\label{fig9}

\bc
\includegraphics[width=8cm]{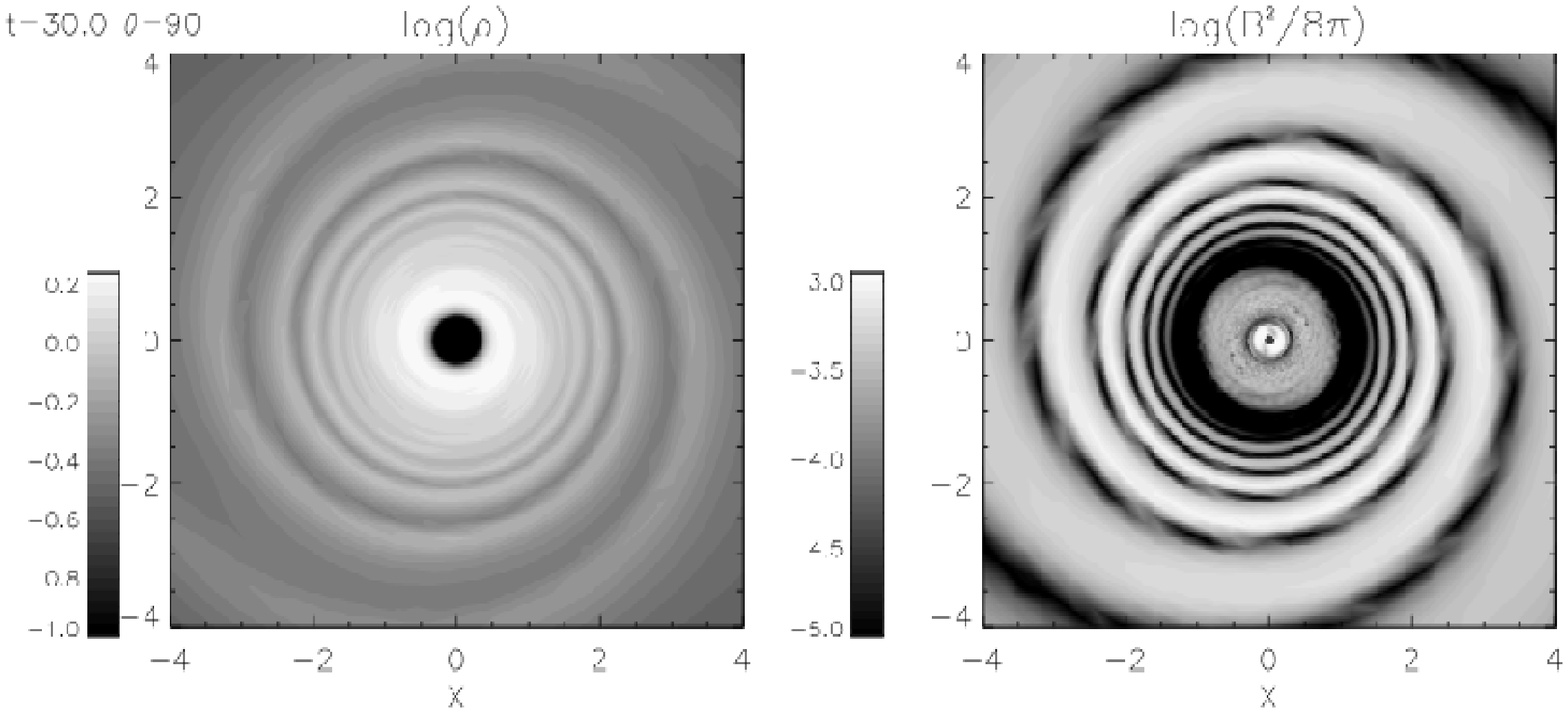}
\ec
\caption{Same as fig.\ref{fig8}, but for the case of $\theta=90\Deg$.}
\label{fig10}

\end{figure}

\subsection{Initially inclined field: V, A and PR fields}

The cases for $\theta=45\Deg$ and 80\deg show that tilted lines of force are wound up to create S configuration near the disk plane.  The field lines penetrating the inner disk are also wound up significantly and their tension force makes the magnetic field "stand up". The twisting magnetic field lines remove the angular momentum of the gas, and the gas is accreted to the central region with the field line. Eventually, due to the dissipation and upward propagation of the tightly wound up toroidal magnetic field in the central region, V field is produced. 

In the case of $\theta=45\Deg$, the ratio of each component of the magnetic field strength is about $|B_r| :|B_\phi| :|B_z| \sim 3:10:1$ in the central region ($0.3<r<0.8$). The ratio is similar to that obtained in the accretion disk in which the magneto-rotational instability is developed. 

The highly inclined field with $\theta=80\Deg$ similarly creates S in the disk and V field in the central region. Thus, outer S and inner V configurations of galactic magnetic fields are created simultaneously from a tilted primordial field, not strongly dependent on the inclination angle.

When the initial field was more vertical, an A type configuration is dominant in the outer disk. It should be noted that the A field direction is reversed with respect to the galactic plane. Such plane-reversed (PR) field is indeed observed in the local disk of the Milky Way.  

\subsection{Initially parallel field and Grand-designed spiral arms: V and S field}
 
The case for $\theta=90\Deg$, e.g. for a primordial field parallel to the disk plane, results in an S field almost over the entire disk. The field intensity map shows that two-armed grand-design spiral pattern is formed as shown in Fig \ref{fig10}, where $B^2/8 \pi$ is shown by grey scale intensity and is compared with the gas density. It must be stressed that the gaseous spiral arms are formed along the minimum loci of the field intensity (neutral sheet). 
The similar density structure is also appeared for various inclination angles except for $\theta=0\Deg$ 
(see Fig. \ref{fig8} and \ref{fig9}) 
In the central region, the field lines are tightly wound, resulting in high magnetic pressure. 
This causes inflation of magnetic flux, and also creates the V field in the central region (Fig. \ref{fig6}(f)).
The ratio of each component of the magnetic field strength is about $|B_r| :|B_\phi| :|B_z| \sim 1:100:10$ in the central region ($0.3<r<0.8$).
The ratio shows that the field structure is almost R field in the central region. 

\subsection{Relation of created fields and the initial field inclination}
 
\begin{figure} \bc   
\includegraphics[width=8cm]{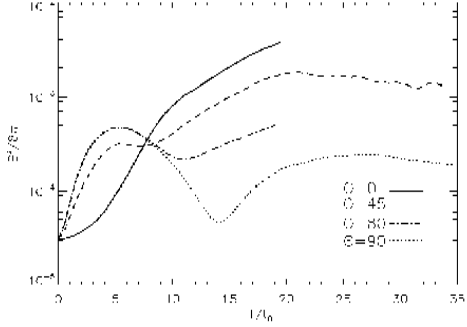}
\ec 
\caption{Time evolution of the magnetic energy density per volume $B^2/{8\pi}$. Initially vertical fields are more strongly amplified 
compared with initially parallel fields.}   
\label{fig11}  
\end{figure}

Figure \ref{fig11} shows the time evolution of the magnetic   energy per volume, $B^2/{8\pi}$, for various inclination of the initial field direction. The magnetic energy density per volume is averaged within the central region of the disk ($0.3 <r< 0.8$, $|z|<0.04$). The field strength for initially vertical field is strongly amplified. On the other hand, if the initial field was parallel to the disk, the azimuthal field is not drastically amplified.  

The strength of the created spiral field is weaker than that for smaller inclination, because the parallel field lines to the disk plane can be easily dissipated when it is tightly wounded. The larger is the inclination (more vertical) of the original field, the stronger is the created spiral field. In other words, the centrally accumulated vertical field is proportional to the total flux of original field penetrating the initial galactic disk. If the disk thickness is ignored, the total flux $F$, and therefore, the final amplified field strength is roughly represented by $F\sim F_0 {\rm cos}\theta$. Namely, the total flux was larger for larger inclination, and vice versa. 

This relation between the initial angle of the primordial field and the resulting field strength predicts that, if the initial field was perfectly parallel to the disk plane, e.g. $\theta=90^\circ$, the resultant field strength would be very weak. Also, the total flux of magnetic field of a galaxy reflects the initial field inclination. We may thus introduce a field-strength function for spiral galaxies depending on the initial inclination. If the initial disk directions of primordial galaxies were randomly distributed with respect to the large-scale cosmic magnetic field, the number density of galaxies with total field flux $F$ would be represented by the probability distribution of angle $\theta$ as $dN(F)/dF=C {\rm sin}\theta=C' \sqrt{1-(F/F_0)^2}$. Instead of $F$ we may use the mean field strength $\bar B$ and $\bar B_0$ as
$dN(B)/dB=C'' \sqrt{1-\bar B/\bar B_0}$.

\begin{figure}
 \bc   
\includegraphics[width=8cm]{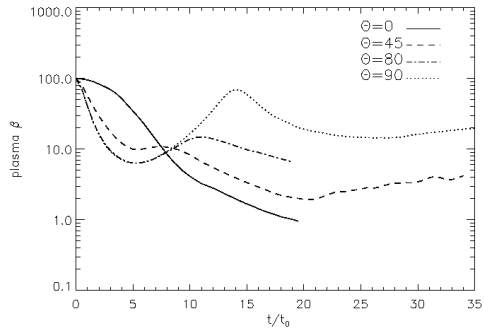}
\ec 
\caption{Time evolution of plasma $\beta$ value, e.g. the ratio of the gaseous to magnetic energy densities. 
The more vertical is the initial field, the stronger is the decrease of $\beta$.} 
\label{fig12} 
\end{figure}

\begin{figure}
\bc   
\includegraphics[width=8cm]{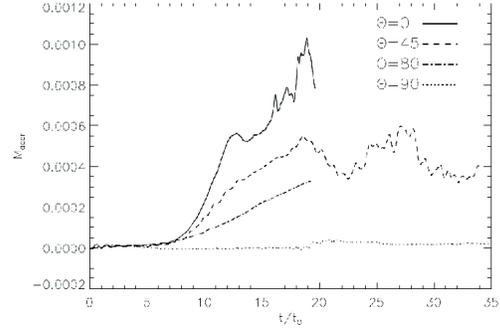}
\ec
\caption{Accretion rate of the disk gas as functions of time. Accretion, and therefore the magnetic braking, is more efficient for initially vertical field compared to the initially parallel field. .}  
\label{fig13}   
\end{figure} 

\subsection{Initial field direction and the magnetization of the ISM}
 
Figure \ref{fig12} shows the time evolution of plasma $\beta$, the ratio of the thermal energy density of gas to magnetic energy density in the disk. The plasma $\beta$ is also averaged value within the central region of the disk ($0.3 <r< 0.8$, $|z|<0.04$). The more vertical was the initial field to the disk, the more strongly decreases the $\beta$. The reason is simply because of the initial magnetic flux penetrating the whole disk, as discussed in the previous subsection related to the azimuthal field amplification.

\subsection{Magnetic accretion of the disk, angular momentum transfer, and jet formation}

Figure \ref{fig13} shows the time variation of accretion rate of gas in the disk for the various inclination of the initial field direction. The accretion of gas takes place due to the angular momentum transfer due to the magnetic tension by twisted vertical field anchored to the intergalactic space through the ${\bf j} \times {\bf B}$ term in equation of motion.  Accordingly, the accretion rate is strongest for the case of initial field perpendicular to the disk, as shown in figure \ref{fig13}, while it is low for parallel field case. 
The accretion flow will enhance the vertical component of the magnetic filed in the very central region of the galaxies ($< 100$ pc), although it is not resolved in the presenting simulations. 

 The twisted field yields magnetic pressure gradient in the direction perpendicular to the disk, and produces a polar cosmic jet. The jet formation is strongest, when the initial field is vertical. This is a well known mechanism to create a cosmic jet since Uchida and Shibata (1985). It should be emphasized that even for a nearly parallel initial field with $\theta \sim 80\Deg$ the accreted vertical component in the galactic center is amplified enough to cause a polar jet. Thus, the vertical field and jet formation near the nuclei are common phenomenon in almost all spiral galaxies, if the present primordial origin hypothesis applies.

\subsection{Field amplification and magneto-rotational instability}

\begin{figure*} 
\begin{center}
\includegraphics[width=15cm]{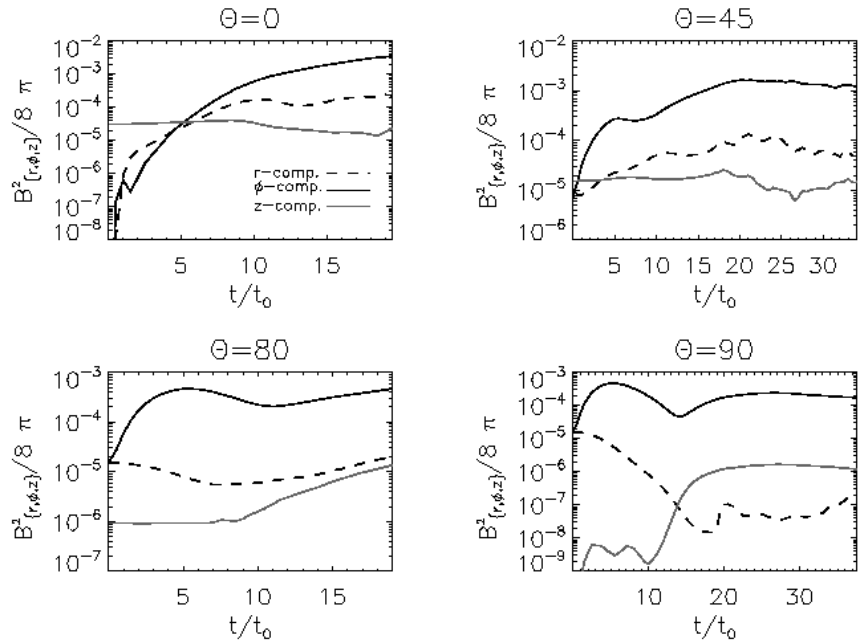} 
\end{center}
\caption{Time evolution of magnetic energy density for individual magnetic components. Thick solid, long-dashed and thin-dotted lines show $B_\phi^2/(8\pi)$, $B_r^2/(8\pi)$, and $B_z^2/(8\pi)$, respectively. The four panels indicate the cases for different inclination angles: $\theta=0^\circ$, $\theta=45^\circ$, $\theta=80^\circ$, and  $\theta=90^\circ$, respectively, from top left to bottom right.}
\label{fig14}
\end{figure*}

The magnetic Reynolds number in a rotating disk is defined by Sano \& Miyama (1999) as $R_{\rm mc} \equiv v_{{\rm A}z}^2/(\eta_0 \Omega)$, and is a good indicator to investigate the magneto-rotational instability (MRI). Here, $v_{{\rm A}z}=B_z/\sqrt{4\pi\rho}$, $\Omega$ is angular frequency of the disk, $\eta_0$ the magnetic diffusivity, and $B_z$ the $z$ component of the magnetic field. When $R_{\rm mc}$ is less than unity, the dissipation process generally suppresses the growth of instability. The magnetic Reynolds number may be expressed in terms of  $R_{\rm m0}$ as $R_{\rm mc}= R_{\rm m0} (c_{\rm s}/v_0)^2 (v_{{\rm A}z}/c_{\rm s})^2$, where $c_{\rm s}$ is the sound speed in the disk. Since the temperature in the disk is assumed to be $\sim 10^4$ K, we have $c_{\rm s}/v_0 \sim H/r_0 \sim 0.05$ in the disk, where $H$ is the scale height of the disk. In the case of $\theta=0^\circ$, we have $(v_{{\rm A}z}/c_{\rm s})^2 = (2/\beta \gamma) \simeq 1/50$ at the initial condition,
  which leads to $R_{\rm mc} \simeq 0.1$. The wavelength for maximum instability is given by $\lambda_{\rm m} \sim \eta_0/v_{{\rm A}z} \sim 0.07 r_0$. This wavelength is slightly larger than the scale height of the disk. Therefore, the axi-symmetric mode of the MRI is suppressed by the magnetic diffusivity in the present case. 

Even when the MRI is suppressed inside the disk, it can develop near the surface of the disk (Sano \& Miyama 1999). Near the disk surface, the wavelength for maximum instability becomes shorter than the scale height, because the Alfv\'en speed is larger for the lower density. Once the MRI develops there, surface accretion occurs even in the resistive disk (Kuwabara et al. 2000, 2005). These processes amplify the strength of the magnetic field. Fig.\ref{fig14} shows the time evolution of the magnetic energy density  for each component of the magnetic field, where the magnetic energy densities are the values averaged within the disk at $0.3 < r < 0.8$ and $|z|<0.04$. Each panel shows the result for different inclination angle of the initial magnetic field.

In the case of $\theta = 45^\circ$, the initial amplification of the
magnetic field (around $t < 5$) is caused by the differential rotation
of the disk and the energy density of $B_\phi$ increases at first. It
begins to increase almost exponentially around after $t \sim 8$ when
the accretion has become prominent (see Fig. \ref{fig13} ). The energy
density of $B_r$ increases almost exponentially from the beginning.
The energy density of $B_z$ also slightly increases, and the ratio of
between the individual magnetic components remains at about
$|B_r|:|B_\phi|:|B_z| \sim 3:10:1$. The ratio is a typical value that
is obtained in an accretion disk with developing MRI (Balbus and Hawley 
1991). We also found that the speed of the accretion flow near
the disk surface is larger than that at $z=0$ where the flow speed is
sometimes positive in the $r$ direction. Thus, we may conclude that
the magneto-rotational instability develops near the surface of the
disk. The total magnetic energy is eventually saturated around $t \sim
20$. In the case of $\theta = 0^\circ$, the time evolution is similar
to that of $\theta = 45^\circ$, although the magnetic energies are not
saturated yet.


The growth rate of the field strength, as represented by the time in which the amplitude increases by a factor of $e$, is found to be approximately equal to $0.1/t_0$ from figure \ref{fig14} and figure \ref{fig11}. This rate may be compared with the growth rate of perturbation as predicted from the linear theory of MRI. The maximum local growth rate predicted from the rotation curve of the Miyamoto-Nagai potential is about $0.7/t_0$ at $0.3<r/r_0<1$ within an ideal MHD approximation. The growth rate as we found here is slightly smaller than this maximum value, probably because the MRI in our simulation is suppressed by the the resistive dissipation process. The result is consistent with the global linear theory by Sano \& Miyama (1999) which studied MRI in a resistive stratified Keplerian disk with vertical magnetic field. The global linear analysis of the surface accretion has not been done yet for our initial setup. The more quantitative comparison with the global linear analysis would be subject for future work.

In the case of $\theta = 90^\circ$, the initial amplification is also caused by the differential rotation of disk, as in the case of $\theta = 45^\circ$. Then, the energy density of $B_\phi$ decreases around $t \sim 10$ because of the magnetic diffusion or by reconnection in the disk. The energy density of $B_r$ also decreases because of the magnetic diffusion or reconnection. A small amount of $B_z$ field is amplified in the disk around $t \sim 13$ possibly because of the vertical motion driven by the thermal pressure during the magnetic diffusion or reconnection. The $B_z$ field is stretched by the rotation of the disk, and $B_\phi$ field is amplified again. Because the accretion flow does not develop in time (see Fig.13), $B_r$ does not increase. In the case of $\theta = 90^\circ$, the MRI near the surface of the disk does not develop because there is no global vertical field crossing the disk.  Each component of magnetic energy eventually saturates around $t \sim 20$ and 
 keeps almost constant during our calculation. In the case of $\theta = 80^\circ$, time evolution is something between that of $\theta = 45^\circ$ and $\theta = 90^\circ$, although the magnetic energies have not saturated yet.

Even though the axi-symmetric mode of the MRI is suppressed in the disk, the non axi-symmetric mode of MRI may develop when a weak toroidal component of the magnetic field is dominated in the disk (e.g., Balbus and Hawley 1992; Terquem \& Papaloizou 1996). However, there is no indication of non axi-symmetric mode of the MRI, even in the case of $\theta = 90^\circ$. The most-unstable wavelength of the non axi-symmetric mode may not be resolved in our simulation, because only 64 grid points are used in the $\phi$ direction. The longer wavelength may develop later, but it may take time because the linear growth rate of the non axi-symmetric MRI is generally smaller than that of the axi-symmetric one (e.g., Matsumoto \& Tajima 1995). The further simulations with fine grid size and longer time scale will be needed in future. 
 
\section{Discussion}
 
We have proposed a new scenario, which we call the primordial-origin model, in order to explain the observed composite magnetic configurations in spiral galaxies. A tilted weak uniform magnetic field is wound up by a rotating disk galaxy and creates the four basic magnetic topologies, S, A, R and V (bisymmetric spiral, axisymmetric spiral, ring, and vertical). We also showed that the reversal of field direction with respect to the galactic plane (PR - plane reversal - configuration), as observed in the local disk of the Galaxy, is formed from the vertical component of primordial field during the formation of A configuration.  In order to show that this scenario indeed applies, we have performed three-dimensional MHD simulations of magnetized rotating gas disk, mimicking a forming disk galaxy.

We found that the more vertical was the initial field, the stronger is the created field. The V field is created and amplified in the central region even from tilted uniform field. The magnetic fields influence the interstellar gas, and the twisted vertical field accelerates outflow of the central gas toward the halo. This leads to angular momentum loss of the disk, and enhances the disk accretion. The S field creates two-armed spiral arm structure, and the gas is compressed along the spiral neutral sheets, yielding a grand-designed spiral structure in the gaseous disk.

We focused on the global topology of galactic magnetic fields, and explained the variety of observed configurations from a unified primordial-origin hypothesis. On the other hand, detailed analyses of the interstellar physics are still crude. In this context our paper may be supplementary to the previous MHD simulations of magnetized disks such as those by Kudoh et al. (1998), Matsumoto (1999), Machida et al. (2000), Nishikori et al. (2006), and more recently by Wang and Abel (2009), in which detailed analyses and discussions are given for the ISM physics.
 
The main purpose of this paper was to propose the idea for explaining the origin of the different global magnetic configurations in spiral galaxies. The idea was basically confirmed by MHD simulations of the global galactic disk with the empirical turbulent magnetic diffusivity. The numerical simulation showed that the vertical (V) structure of magnetic field in the central region of the galaxy is created, even when the initial inclination angle of the magnetic field is nearly parallel to the galactic disk. It also showed that the S,  A, or PR magnetic field topologies are realized depending on the initial conditions. The magnetic energy density was found to be saturated after several rotations in the central region of the disk. The averaged amplitude of the saturated magnetic field in the central region ($<$ 1kpc) was about $\sim 7 \mu$G in the case of $\theta =45\Deg$, and $\sim 2 \mu$G in the case of $\theta =90\Deg$. 

The amplification of the global magnetic field is caused by the differential rotation of the disk as well as the accretion flow to the central region. The field strength gets saturated by the turbulent diffusivity after some rotations in the central disk. However, in order to discuss the strength of the magnetic field in the solar neighborhood, we need further simulations with longer time scales. The galactic dynamo process may be needed to get further amplification of the magnetic field (e.g., Hanasz et al. 2009).  We have also considered possible growth of the magneto-rotatational instability (MRI). The instability is suppressed inside the galactic disk around the galactic plane. However, it was found that the MRI grows near the surface of the disk, and magnetic fields are also amplified by MRI.

In the MHD simulation, we have assumed an initially uniform magnetic field as the primordial seed field. This uniform assumption may apply in such cases that the uniformity of the field was sufficiently larger than the scale radius of proto-galactic disk. Namely, the present scenario and the MHD simulation may be applied to such circumstances that intergalactic magnetic fields existed with scale lengths comparable to the size of a cluster of galaxies. 
Since the major interest was to see the topological evolution of initially uniform magnetic field in a single galaxy, the calculated area covered only a few tens of kpc around the galaxy. Moreover, the initial field was assumed to be already as strong as $\sim 0.6 \mu$G. In this context, the present simulation was not aimed at fully tracing the cosmological galaxy formation, starting from much smaller density perturbations and weaker intergalactic field before structural formation in the universe.

In order to see if a much weaker primordial magnetic field can be amplified to a micro G level during the galaxy formation, we here try an order-of-magnitude estimate of possible evolution of the field strength along the idea of primordial origin hypothesis (see also Sofue and Fujimoto (1987)).
Suppose that an intergalactic primordial magnetic field of one nG was trapped to a pre-galactic sphere of $r_{\rm p} \sim 100$ kpc radius, and that the sphere collapsed to a proto-galactic disk of $r_{\rm d} \sim 10$ kpc. Then the field strength is amplified by a factor of $(r_{\rm d}/r_{\rm p})^2 \sim 100$, yielding $0.1 \mu$G. If a half of the disk flux was further accumulated to the central region, forming a proto bulge before star formation, of radius $r_{\rm b} \sim 1$ kpc, it is further amplified by $(r_{\rm d}/r_{\rm b})^{2} \sim 100$ times. Thus, amplification by a factor of $100\times 100=10^4$ times may easily be attained to create a central disk field of $\sim 10 \mu$G. Moreover, the twist by the galactic rotation, as well as accretion by magnetic braking, would further amplify the vertical field by a factor of 10 to 100 times, creating vertical fields of a few mG field. Thus, we may have $\sim 10^6$ times amplification to a few $\mu G$ in the disk, and further $\sim 10^6$ times amplification to mG field in the center, starting from intergalactic primordial field on the order of $\sim 1$nG.

Finally we comment that the self gravity, star formation, and more sophisticated ISM conditions such as the existence of multi-phase components and their phase transitions in a more realistic galaxy-formation model should be subject for future works. The present simulation is, therefore, preliminary and should be taken as the initial-attack toward more general studies of the origin of galactic magnetism in the scheme of galaxy formation, considering the self-gravity and cosmological magnetic field origin.

\vskip 3mm
Numerical calculations were carried out on NEC SX-9 at the Center for Computational Astrophysics (CfCA) of National Astronomical Observatory of Japan

\end{document}